\documentclass[onecolumn,10pt]{article}
\usepackage{authblk}
\usepackage{amsmath}
\usepackage{amsfonts}
\usepackage{amssymb}
\usepackage{amsthm}
\usepackage{times}
\usepackage{cancel}
\usepackage{dsfont}
\usepackage{graphics}
\usepackage{makeidx}
\makeindex

\newcommand{\ket}[1]{| #1 \rangle}
\newcommand{\bra}[1]{\langle #1 |}

\begin{document}

\date{}
\title{There is No Quantum World\footnote{Forthcoming in Jan Faye and Lars-Göran Johansson, \emph{How to Understand Quantum Mechanics: 100 years
of Ongoing Interpretations, Boston Studies in Philosophy and History of Science}, Springer.}}
\author{Jeffrey Bub} 
\affil{Philosophy Department,
Institute for Physical Science and Technology, Joint Center for Quantum Information and Computer Science,		
University of Maryland}
\maketitle

\begin{abstract}
I outline a neo-Bohrian interpretation of quantum mechanics---a view of quantum mechanics that accords with the core insights in Bohr's thinking, with a twist that justifies the prefix `neo.' In a second part of the paper, I show how von Neumann's work on infinite direct products provides a theoretical framework that deflates the measurement problem and justifies Bohr's  insistence on the primacy of classical concepts.
\end{abstract}

%\printindex

According to Aage Petersen \cite[p. 12]{Petersen} who was Bohr's  assistant from 1952 to 1962, when Bohr was asked whether quantum mechanics could be understood as describing or mirroring an underlying quantum world, his response was 
\begin{quote} There is no quantum world. There is only an abstract quantum physical description. It is wrong to think that the task of physics is to find out how nature \emph{is}. Physics concerns what we can say about nature.'
\end{quote}  A similar statement does not appear in any of Bohr's  published writing, so it is not surprising that the attribution to Bohr  has been derided (see \cite{Mermin}). But there is no reason to doubt Petersen and his recollection of Bohr's  response, properly understood, is the key to understanding how Bohr  saw the transition from classical to quantum mechanics.

This paper is divided into two parts. In the first part I present a neo-Bohrian\index{neo-Bohrian} interpretation of quantum mechanics---a view of quantum mechanics that accords with the core insights in Bohr's  thinking, with a twist that justifies the prefix `neo.' The discussion here follows closely my earlier writing on this topic \cite{Bub_Foreword, Bub_HSNS}. The second part deals with the measurement problem\index{measurement problem}, specifically how to put away
the measurement problem\index{measurement problem}. 

In a 1935 letter, John von Neumann wrote to Garrett Birkhoff \cite{Neumann_letter}
\begin{quote}
I would like to make a confession which may seem immoral: I do not believe absolutely in Hilbert space any more. After all Hilbert-space (as far as quantum-mechanical things are concerned) was obtained by generalizing Euclidean space, footing on the principle of `conserving the validity of all formal rules' \ldots  Thus Hilbert-space is the straightforward generalization of Euclidean space, if one considers the vectors as the essential notions. Now we begin to believe that it is not the \emph{vectors} which matter but the lattice of all linear (closed) subspaces. Because: 1) The vectors ought to represent the physical \emph{states}, but they do it redundantly, up to a complex factor, only 2) and besides, the states are merely a derived notion, the primitive (phenomenologically given) notion being the qualities which correspond to the \emph{linear closed subspaces}. But if we wish to generalize the lattice of all linear closed subspaces from a Euclidean space to infinitely many dimensions, then one does not obtain Hilbert space, but that configuration which Murray and I called `case $\Pi_1$.' (The lattice of all linear closed subspaces of Hilbert space is our `case I$_\infty$.')
\end{quote}
I argued in earlier work \cite{Bub_infinite1, Bub_infinite2, Bub_infinite3}   that von Neumann's consideration of infinite direct products in \cite{Neumann_infinite}  provides a framework for resolving the measurement problem\index{measurement problem}. In a recent series of articles \cite{Bossche+1, Bossche+2, Bossche+3, Bossche+4}, Van Den Bossche and Grangier  provide an exposition of von Neumann's  result along similar lines but in considerably more depth.  I sketch the core idea following Van Den Bossche and Grangier but, as I see it now, since no physical system is literally infinite in this sense, von Neumann's generalization of Hilbert space really shows  that the measurement problem\index{measurement problem} is a feature of quantum mechanics, not a bug, as Michael Cuffaro aptly puts it \cite{Cuffaro_feature}, and this is consistent with Bohr's  view.

\section{A Neo-Bohrian\index{neo-Bohrian} Interpretation of Quantum Mechanics}

 Quantum mechanics began with the publication in 1925 of Werner Heisenberg's breakthrough paper   `On the quantum-theoretical re-interpretation (`Umdeutung') of kinematical\index{kinematic} and mechanical relations' \cite{Heisenberg}. As Heisenberg saw it, the quantized orbital paths of electrons revolving around a central nucleus in Bohr's  theory of the atom, which conflicted with classical electrodynamics as well as classical mechanics, were unphysical. To get rid of  the orbits, he proposed  `a theoretical quantum mechanics \ldots in which only relations between observable quantities occur.' He accomplished this by the extraordinary manoeuvre of `re-interpreting' classical mechanical quantities like position and momentum as  operators represented by arrays, which Max Born identified as matrices when Heisenberg sent him a draft of the paper. The result of applying operations represented by the operators $A$ and $B$ can differ from the result of applying the operations in the reverse order, which is to say that the re-interpreted matrix quantities needn't commute. 

 Erwin Schr\"{o}dinger published a wave-mechanical version of the theory in 1926 that kept the orbits and explained their quantization as a wave phenomenon. He proved the empirical equivalence of the two versions  of quantum mechanics for experiments relevant at the time. The general theoretical question of equivalence was definitively settled positively in 1927  in John von Neumann's reformulation of quantum mechanics as a theory of  `observables' represented by operators and states represented by rays in Hilbert space \cite{Neumann27}, \cite{Duncan}. 
 
In 1935 Einstein, Podolsky, and Rosen \cite{Einstein+} drew attention to the existence of non-separable or `entangled' states\index{entanglement!entangled states!} of separated systems  and exploited their correlations to argue for the incompleteness of quantum mechanics. John Bell re-examined the  argument in 1964  in a seminal article \cite{Bell} in which he proved that any deterministic extension of quantum mechanics with additional `hidden' variables would necessarily introduce what Einstein \cite{Einstein_spooky}   called `spooky action at a distance' in the theory, in the sense that the outcome of a measurement could be influenced instantaneously by the setting of a remote measuring instrument. For Schr\"{o}\-dinger,  entanglement\index{entanglement} was `\emph{the} characteristic trait of quantum mechanics, the one that enforces its entire departure from classical lines of thought' \cite[p. 555]{Schrodinger_entangle}. Following Bell's proof, the probabilistic correlations of entangled quantum states are recognized as crucial to applications like quantum computation, quantum cryptography, and quantum communication.
  
 Heisenberg's `reinterpretation' of classical quantities like position and momentum as noncommutative replaces the commutative algebra of physical quantities of a classical system with a noncommutative algebra of `observables.' To understand  what noncommutativity involves it is helpful to think of two-valued observables. These represent measurable properties of a quantum system or the corresponding propositions, for example, the property  that the energy of the system lies in a certain range, with the two values of the observable representing `yes' or `no,' or equivalently the proposition asserting that the energy lies in a certain range, with the two values of the observable representing `true' or `false.' The two-valued observables of a classical system form a Boolean algebra\index{Boolean algebra}. 
 
 We all, in a loose sense, understand the concept of a Boolean algebra\index{Boolean algebra}, even if the term is unfamiliar. A Boolean  algebra is simply a formalization of the way  we picture properties or propositions fitting together when we combine them with conjunction, disjunction, negation, the logical connectives `and,' `or,' `not,' as a classical or commonsense `state of reality'  in which every proposition is assigned a truth value, either `true' or `false,' consistently with the connectives. George Boole characterized a Boolean algebra\index{Boolean algebra} as capturing `the conditions of possible experience' \cite[p. 229]{Boole},\cite{Pitowsky}. He added: `When satisfied they indicate that the data \emph{may} have, when not satisfied they indicate that the data \emph{cannot} have, resulted from actual observation.' A Boolean algebra\index{Boolean algebra} is isomorphic to a set of subsets of a set. The algebraic operations correspond to the intersection, union, and complement of a set, equivalently to conjunction, disjunction, and negation. 
 
 Quantum mechanics replaces  the Boolean algebra of subsets of classical phase space, where the points represent classical states and subsets represent ranges of values of dynamical variables, with a non-Boolean algebra\index{Boolean algebra}, formalized as the algebra of closed subspaces of a Hilbert space, a vector space over the complex numbers, or equivalently a projective geometry. So the transition from classical to standard quantum mechanics is, formally, the transition from a Boolean algebra\index{Boolean algebra} of subsets of a set to a non-Boolean algebra\index{Boolean algebra} of subspaces of a vector space.\footnote{For an illuminating account of the significance of non-Booleanity, see Janas, Cuffaro, and Janssen \cite{Janas+}. They show, for a particular case, that that classical probabilistic correlations, represented geometrically, are constrained to lie in a tetrahedron, but quantum correlations can lie outside the tetrahedron and are bounded by an elliptope (roughly, a `fat' tetrahedron).}

Classical theories are Boolean theories. To say that the algebra of observables of a quantum system is noncommutative is equivalent to saying that  the sub-algebra of properties or propositions is non-Boolean. The non-Boolean algebra of quantum mechanics is formally a collection of  Boolean algebras\index{Boolean algebra} that are `intertwined,' as Gleason put it \cite[p. 886]{Gleason}, in such a way that the whole collection can't be embedded into a single inclusive Boolean algebra\index{Boolean algebra}, so you can't assign truth-values consistently to the propositions about observable values in all these Boolean algebras\index{Boolean algebra}. In fact, there are finite sets of Boolean algebras\index{Boolean algebra} in the collection of Boolean algebras\index{Boolean algebra} of a quantum system that don't fit together into a single Boolean algebra\index{Boolean algebra}, unlike the corresponding collection for a classical system. Kochen and Specker, who first proved non-embeddability \cite{Kochen+}, call this structure a `partial Boolean algebra.'\index{Boolean algebra}. 

Bohr  did not refer to Boolean algebras\index{Boolean algebra}, but the concept is simply a precise way of codifying the salient feature of Bohr's  interpretation of quantum mechanics: his insistence that `the account of all evidence must be expressed in classical terms,' that's to say `unambiguous language with suitable application of the terminology of classical physics,' for the simple reason that we need to be able `to tell others what we have done and what we have learned' \cite{Bohr2}. The significance of `classical' here as being able `to tell others what we have done and what we have learned' is that the events in question should fit together as a Boolean algebra\index{Boolean algebra}, so conforming to Boole's `conditions of possible experience.' A `world' in which it is possible to talk about truth and falsity, referring to \emph{this} rather than \emph{that}, is Boolean.  In this sense, \emph{there is no noncommutative or non-Boolean quantum `world'}---no way to fit together the properties of a quantum system into a complete catalog, no consistent assignment of truth values to the corresponding propositions. Moreover, from this perspective, since quantum mechanics does not represent physical reality in the realist sense, the theory functions rather as a tool that tells us `what we can say' about the physical world. As Cuffaro points out \cite{Cuffaro_say}, a  quantum state represents `the structure of and interdependencies among the possible ways that one can effectively characterize a system in the context of a physical interaction.'

In a quantum theory, the Boolean algebra\index{Boolean algebra} of a classical theory is replaced by a collection of Boolean algebras\index{Boolean algebra}, representing different Boolean perspectives or Boolean frames associated with different complete commuting sets of observables yielding incompatible measurement outcomes. Because there is no consistent assignment of values to the observables of a quantum system, quantum probabilities can't refer to uncertainty about possible but unknown values of observables. Rather, quantum probabilities are, as von Neumann pointed out, `perfectly new and \emph{sui generis} aspects of physical reality' \cite{Neumann_sui}  about the likelihood of getting a certain random outcome i\emph{n a measurement of an observable}. A measurement is associated with the selection of a particular Boolean frame in the family of Boolean algebras\index{Boolean algebra} associated with a quantum system. In terms of observables,  a measurement involves the selection of an orthogonal basis defined by a set of commuting observables in Hilbert space. As a consequence, the observer is no longer `detached,' unlike the observer in classical mechanics, as Pauli observed in a letter to Max Born \cite[p. 218]{Born}. 
 
Quantum probabilities are `uniquely given from the start,' in the sense von Neumann explained  in an address  to an international congress of mathematicians in Amsterdam, September 2--9, 1954. Here is the relevant passage \cite{Redei+}:
\begin{quotation}
Essentially if a state of a system is given by one vector, the transition probability in another state is the inner product of the two which is the square  of the cosine of the angle between them.\footnote{Von Neumann evidently meant to say that the transition probability is the  square of the (absolute value of) the inner product, which is the square  of the cosine of the angle between them.}  In other words, probability corresponds precisely to introducing the angles geometrically. Furthermore, there is only one way to introduce it. The more so because in the quantum mechanical machinery the negation of a statement, so the negation of a statement which is represented by a linear set of vectors, corresponds to the orthogonal complement of this linear space. And therefore, as soon as you have introduced into the projective geometry the ordinary machinery of logics, you must have introduced the concept of orthogonality. This actually is rigorously true and any axiomatic elaboration of the subject bears it out. So in order to have logics you need in this set a projective geometry with a concept of orthogonality in it. 

In order to have probability all you need is a concept of all angles, I mean angles other than $90^{\circ}$. Now it is perfectly quite true that in geometry, as soon as you can define the right angle, you can define all angles. Another way to put it is that if you take the case of an orthogonal space, those mappings of this space on itself, which leave orthogonality intact, leave all angles intact, in other words, in those systems which can be used as models of the logical background for quantum theory, it is true that as soon as all the ordinary concepts of logic are fixed under some isomorphic transformation, all of probability theory is already fixed. 

What I now say is not more profound than saying that the concept of a priori probability in quantum mechanics is uniquely given from the start. 
\end{quotation}

If relativity is about space and time, quantum mechanics is \emph{about probability}, in the sense that quantum probabilities are `\emph{sui generis}' and `uniquely given from the start' as an aspect of the kinematic\index{kinematic} structure of the theory and are not imposed from outside as a measure of ignorance, as in classical theories, where probability is a measure over phase space (so a measure over classical states). In this non-Boolean framework, new sorts of nonlocal probabilistic correlations associated with entanglement\index{entanglement} are possible, which makes quantum information fundamentally different from classical information. In a Boolean theory such correlations are impossible without introducing `spooky' action at a distance (see \cite{Bub_book}). What's puzzling, from a Boolean perspective, is that measurement in a non-Boolean theory is not passive---not just `looking' and registering what's there in a passive sense. A measurement requires the selection of a basis of commuting observables and produces an intrinsically random change in the description accompanied by a loss of information from prior measurements about observables not commuting with the measured observable, and that's not how we are used to thinking of measurement in a Boolean theory where any change due to measurement is non-random and can always be accounted for in principle. 

To sum up: In the non-Boolean theory of quantum mechanics, probabilities arise via  the Born rule as a feature of the geometry of Hilbert space. These probabilities can't be understood as quantifying ignorance about the pre-measurement value of an observable, as in a Boolean theory. Rather, they represent a new sort of ignorance about something that doesn't yet have a truth value, something that simply isn't one way or the other before we measure, something that requires us to act and do something that we call a measurement before nature supplies a truth value---and changes the truth values of incompatible propositions that don't belong to the same Boolean frame, associated with observables that don't commute with the measured observable. On a neo-Bohrian\index{neo-Bohrian} view, quantum mechanics is a new sort of non-representational theory for an irreducibly indeterministic universe, with a new type of nonlocal probabilistic correlation  for entangled quantum states  of separated systems, where the correlated events are  intrinsically random, not merely apparently random like coin tosses. The `neo' here is added to indicate the interpretation of  Bohr's use of `classical' as `Boolean.'

Hilbert space provides the kinematic\index{kinematic} framework for such an irreducibly indeterministic universe in a similar sense to which Minkowski space provides the kinematic\index{kinematic} framework for the physics of  a non-Newtonian, relativistic universe. In special relativity, Lorentz contraction is  a kinematic\index{kinematic} consequence of the spatio-temporal constraints on events imposed by the geometry of Minkowski space. In quantum mechanics, the irreducible loss of information in a quantum measurement---Bohr's `finite  and uncontrollable interaction between the objects and the measuring instruments' \cite[p. 700]{Bohr1}---is  a kinematic\index{kinematic} (i.e., pre-dynamic)  consequence of \emph{any} process of gaining information of the relevant sort, irrespective of the dynamical processes involved in the measurement process,  given the objective probabilistic constraints on correlations between events imposed by the  geometry of Hilbert space. 

On this view, quantum mechanics does not provide a representational  explanation of events. Noncomutativity or non-Booleanity makes quantum mechanics quite unlike any theory we have dealt with before, and  there is no reason, apart from tradition, to assume that the theory should provide   the sort of  realist explanation we are familiar with in a  theory that is commutative or  Boolean at the fundamental level.

\section{The Measurement Problem}\index{measurement problem}

There are different formulations of the measurement problem\index{measurement problem}, but the classic formulation is von Neumann's statement in \cite[p. 351]{Neumann_book}. Von Neumann points to two fundamentally different ways in which the state of a quantum system can change: the change that occurs when a system undergoes measurement (he calls this Process \textbf{1}), and the change when a system evolves dynamically under the action of an energy operator $H$ without measurement (Process \textbf{2}). When a system evolves dynamically without being measured, the state $\ket{\psi}$ undergoes a continuous, reversible, unitary transformation in time $t$ to the state
\begin{equation}
\ket{\psi'} = e^{-\frac{2\pi i}{h} tH} \ket{\psi}
\end{equation}
Equivalently, the pure state density operator $W = P_{\ket{\psi}}$, the projection operator onto the state $\ket{\psi}$,  transforms as
\begin{equation}
W \rightarrow W' = P_{\ket{\psi'}}
\end{equation}
But when a system is measured, the state $\ket{\psi}$ undergoes a discontinuous, irreversible, non-unitary transformation  to one of the eigenstates $\ket{\phi_n}$ of the measured observable with probability $|\langle\phi_n|\psi\rangle|^2$. That is, the transition (for a non-selective measurement) is from the pure state $W =P_{\ket{\psi}}$  to a mixed state
\begin{equation}
W \rightarrow W' = \sum_n |\langle\phi_n|\psi\rangle|^2 P_{\ket{\phi_n}}
\end{equation}

For a physical system a measurement is a dynamical interaction with a second system that we characterize as a measuring instrument. The fact that we extract information about the measured system from the interaction has something to do with us, not the measured system. There should be no change in the nature of the dynamical behavior of the measured system just because we choose to regard the interaction as a measurement. The problem is that quantum mechanics requires a dual dynamics, and this is unacceptable without further explanation.

Schr\"{o}dinger calls the measurement problem\index{measurement problem} `the most difficult and most interesting point of the theory' \cite[p. 826]{Schrodinger}. Here's how he put it 
\begin{quote}
(1) The discontinuity of the expectation-catalog {[}the quantum pure state{]} due to measurement is \emph{unavoidable}, for if measurement is to retain any meaning at all then the \emph{measured value}, from a good measurement, \emph{must} obtain. (2) The discontinuous change is certainly \emph{not} governed by the otherwise valid causal law, since it depends on the measured value, which is not predetermined. (3) The change also definitely includes (because of `maximality' {[}the `completeness' of the quantum pure state{]}) some \emph{loss} of knowledge, but knowledge cannot be lost, and so the object \emph{must} change---\emph{both} along with the discontinuous changes and \emph{also}, during these changes, in an unforeseen, \emph{different} way.
\end{quote}

What would count as a solution to this problem? Firstly, we need to hold onto the fact that a measurement is a dynamical interaction between a measured system and a measuring instrument. Appeals to the consciousness of an observer who registers the measurement outcome can't be invoked as a solution. The source of the problem is the noncommutativity or non-Booleanity of quantum mechanics, so we need to look to physical theory for a solution. Secondly, a salient feature of a quantum measurement, say of the spin of an electron, or the polarization of a photon, or the energy of a neutron, is that it involves a macroscopic setup in which we manipulate objects in our Boolean world to extract information about a microsystem, objects such as lasers, ion traps, optical tweezers, polarizing beamsplitters, photodetectors, interferometers, spectrometers, amplifiers, coincidence counters, and the like. The obvious solution would be a demonstration, if possible, that von Neumann's Process \textbf{1} can be reduced to the unitary dynamics of Process \textbf{2}, taking into account that any quantum measurement necessarily involves such systems. 

Von Neumann's work on infinite direct products of Hilbert spaces \cite{Neumann_infinite} (and later work with Murray on operator algebras \cite{Murray}) is relevant here. I sketch the core idea following the  account in Van Den Bossche and Grangier  \cite{Bossche+1, Bossche+2, Bossche+3, Bossche+4}. See also Landsman \cite{Landsman1, Landsman2} and Butterfield \cite{Butterfield}.

Consider, for example, a macrosystem composed of $N$ elementary 2-state systems, say spin-1/2 systems. The Hilbert space representation of the spin algebra for finite $N$ is irreducible: it contains no proper subspace whose vectors remain in the subspace after being acted on by a spin operator, and all representations of the spin algebra are unitarily equivalent to the Pauli representation  (von Neumann's uniqueness theorem \cite{Neumann31}). As $N \rightarrow \infty$ the number of dimensions  $2^N$  becomes uncountable: $2^{N} \rightarrow 2^{\aleph_0} = \mathfrak c= |\mathbb R|$ and the uniqueness theorem no longer holds.\footnote{Thanks to Krzysztof Sienicki for pointing out a correction to my formulation of this point in an an earlier version.}  A Hilbert space with an uncountable number of dimensions is non-separable and decomposes into into a direct sum of $\infty$-dimensional Hilbert spaces, each of which provides a different irreducible representation of the algebra of observables. These inequivalent representations  correspond to different macrostates of the system, where a macrostate is identified with an equivalence class or sector of microstates. A macro-observable takes the same value for every microstate in a given sector. For example, the spin density of an infinite array of spin-1/2 systems defines a macro-observable, the polarization of the macrosystem, that takes different values for different sectors. Dynamical evolutions in which only a finite number of elementary systems undergo a change of state leave the macrostate in the same sector. This sectorization of Hilbert space associated with inequivalent representations of the spin algebra yields a restriction on the superposition principle, a decomposition of the Hilbert space into superselection sectors.

The first phase of a measurement, say a measurement of the polarization of a photon, involves a suitable unitary interaction between the measured system  $S$ and a polarizing beamsplitter, a macroscopic measuring instrument $M$, that leads to an entangled state of the two systems, a linear superposition of product states $\ket{s_i}\ket{m_i} = \ket{s_i,m_i}$ of $S$ and $M$, where the states $\ket{s_i}$ are the  measured polarization states of the photon $S$ and the states  $\ket{m_i}$ are the correlated  states of the beamsplitter $M$.  If the initial states of $S$ and $M$ are $\ket{s} =\sum_i c_i\ket{s_i}$ and $\ket{m_o}$, the transition is
\begin{equation}
\ket{s}\ket{m_o} \rightarrow \sum_i c_i\ket{s_i,m_i}
\end{equation}
The density operator is
\begin{equation}
W_{S,M} = \sum_{ij}c_ic_j^*\ket{s_i,m_i}\bra{s_j,m_j}
\end{equation}
This phase selects a particular Boolean algebra\index{Boolean algebra} in the partial Boolean algebra\index{Boolean algebra} of the composite system (the Boolean algebra generated by the eigenstates of the measured observable and the correlated states of the measuring instrument) and is reversible. 

The second phase involves a macroscopic detector, $D$, such as an avalanche photodiode (a type of photodetector), in which an exponentially increasing  number of elementary systems becomes entangled with the $S$ and $M$ systems in a chain reaction that is not reversible. The density matrix of the composite system becomes almost block-diagonal for large $N$ and, as $N \rightarrow \infty$, the off-diagonal terms vanish in the limit and there is zero interference between sectors. At the infinite limit the vector state of the composite system, a sum of terms from different sectors, represents a mixture---the density operator can be represented as a sum of projection operators onto composite system states, with coefficients representing the probabilities of these states:
\begin{equation}
W_{S,M,D} = \sum_i |c_i|^2\ket{s_i,m_i,d_i}\bra{s_i,m_i,d_i}
\end{equation}
So, at the infinite limit, the probabilities  refer to possible but unknown values of observables.

What's new about the transition to infinity  is that  \emph{vector states need not be pure states on the algebra generated by sectorization}. So, in effect, at the infinite limit, because there is no interference between states belonging to different sectors, states represented by superpositions of vectors from different sectors represent mixtures. Vector states representing superpositions of  states from the same sector do not represent mixtures. 

These considerations show that the measurement problem\index{measurement problem} is resolved at the infinite limit. But infinity never comes. For macrosystems with very large but finite $N$, we can say only that interference between different macrostates  is very small, less than any small number $\epsilon(N)$, and gets smaller and smaller, and harder and harder to detect, as the composite system includes more and more elementary systems. The approximate sectorization of Hilbert space as the number of elementary particles in a system increases without limit, and the consequent negligible interference between different macrostates, explains the  classicality of the macroworld for all practical purposes---FAPP\index{FAPP} to use Bell's acronym---and we can see why we experience the macroworld as classical or Boolean. This  is informative, but it does not explain the occurrence of a definite outcome in a quantum measurement process, so it does not explain how  von Neumann's Process \textbf{1} is reducible to Process \textbf{2}. 

Van Den Bossche and Grangier are clear about this in  \cite[p. 7]{Bossche+1}:  `In principle the radical change of behaviour occurs in the $N \rightarrow \infty$ limit only \ldots.' But in \cite[p. 5]{Bossche+2}, they muddy the issue by saying that `even before reaching the $N \rightarrow \infty$ limit, the sectorisation behaviour sets in and converts the pure state in an \emph{effective} mixed state,’  and `\emph{at some point}, the number of electrons fed into the avalanche is \emph{so large} that it results in a macroscopic change that cannot be ignored.’ (My emphasis.) The crucial words here are `effective' and `at some point' and `so large.' 

What does `effective' mean here, and at what point is `so large’ large enough?  In a measurement of the spin of an electron what we  want is that at the end of the day, so to speak, the avalanche produces a measurement outcome indicating  an electron with positive spin, or an electron with negative spin, with a certain probability. But there is no end of the day in the finite case. All we have is a growing avalanche characterized by an entangled quantum state. The state never `collapses’ to one of the components in the entangled state, so the theory doesn't account for the fact that a measurement outcome actually occurs. It doesn’t matter how many electrons are involved in the avalanche, there is no index in the entangled pure state that indicates a macroscopic change. To say that at some point the pure state is converted to an effective mixed state, or a mixed state FAPP\index{FAPP}, really amounts to saying that at some point, when the entangled pure state is `sufficiently' complex, it is legitimate to ignore the fact that it is neither true nor false that the spin is positive, and  neither true nor false that it is negative, and simply declare that one of these statements is true and the other is false, and interpret the Born probabilities as ignorance probabilities about which of these statements is true, because we lack the technical ability to show otherwise.\footnote{Thanks to Matt Leifer for correcting a similar unjustified claim on my part at a philosophy of physics conference on my book \emph{Bananaworld} \cite{Bub_book} at the University of Western Ontario in 2016.}

Negligible interference is not zero interference. There is always some measurement that could, in principle, reveal an interference effect, which  is to say that the Hilbert space of a composite system composed of $N$ constituent elementary systems does not sectorize for any finite $N$ and von Neumann's problem remains. There is still no explanation about how something that is indeterminate or indefinite can become definite in a measurement process, how something that is neither true nor false in the quantum state of the measured system can become true or false in a measurement, so that the Born probabilities that initially refer to indefiniteness can be understood as probabilities about our ignorance of what is actually the case. 

Nevertheless, von Neumann's result tells us something significant. Granted, there is no solution to the measurement problem\index{measurement problem} in the framework of standard quantum mechanics---the measurement problem\index{measurement problem} is a feature of quantum mechanics, not a bug, to repeat Cuffaro's insightful comment, a necessary consequence of non-Booleanity. There is, however, a solution outside quantum mechanics at the $N \rightarrow \infty$ limit, where the properties of macrosystems modeled as composite physical systems depend on the collective behavior of their elementary constituents and are insensitive to adding or removing any finite number of elementary systems. But since the limit is never reached for any real system composed of a finite number of elementary systems,  there is no number $N$ such that  the partial Boolean algebra\index{Boolean algebra} of a composite quantum system with more than $N$ constituent elementary systems becomes Boolean. So the Boolean macro\-world is not reducible to the non-Boolean quantum level in the usual sense.

Where does this leave things? We know that something happens in a quantum measurement---we see an event occurring, the registration of a definite outcome at the macrolevel. A non-Boolean theory cannot explain how this happens. Yes, interference will be smaller than any designated number $\epsilon(N)$ for large $N$, but for any $N$, no matter how large, there is some interference experiment that could in principle be performed (granted, with practically impossible difficulty) that would reveal interference inconsistent with the measured observable actually taking a definite value---one could derive a contradiction from the assumption that the measured observable takes a definite value and the existence of interference, however negligible. 

So, the occurrence of a definite measurement outcome event must be outside the quantum description. A measurement process can be analyzed to any level of precision in a quantum theoretical treatment, but some `ultimate' macroscopic detecting instrument must always be left out of the non-Boolean analysis as a classical/Boolean reference system for the Born probabilities of the quantum state, which are interpreted as referring to possible measurement outcomes indicated by the `pointer readings' of the detector, one of which actually occurs. The detector itself could, in principle, be analyzed as a finite quantum mechanical system, with the Born probabilities then referring to a classical/Boolean super-detector. While the requirement to leave some part of an experimental set-up outside the quantum description cannot be avoided, the movability of the ``Heisenberg cut'  between the system described quantum mechanically and the macrosystem acting as the detecting instrument is what justifies the account as complete and objective, in the sense that anything relevant to the measurement can potentially be included in the analysis. As Peres put it \cite{Peres+}, `although it can describe anything, a quantum description cannot include everything.' In this sense of `successive' reduction, in which a larger and larger part of an experimental set-up can be treated quantum mechanically, without any limit in principle, the Boolean macro\-world is reducible to the non-Boolean quantum level. On this neo-Bohrian\index{neo-Bohrian} interpretation of quantum mechanics the measurement problem\index{measurement problem} is not resolved but deflated as a feature of non-Booleanity.  
 
Three quotations cited above encapsulate the neo-Bohrian\index{neo-Bohrian} view outlined here:
\begin{enumerate}
\item `There is no quantum world.'  Noncommutativity is formally the replacement of the  Boolean algebra\index{Boolean algebra} of classical mechanics with a collection of intertwined Boolean algebras\index{Boolean algebra} non-embeddable into a single Boolean algebra. Because of non-embeddability the transition from classical to quantum mechanics can't be understood conceptually as replacing  the classical world described by classical mechanics with a quantum world described by quantum mechanics. A non-Boolean physical theory is about `what we can say' about the physical world, not about ontology (see Cuffaro's distinction between `metaphysical realism' and `methodological realism' in \cite{Cuffaro_say} for an elaboration).
\item `The measurement problem\index{measurement problem} is a feature of quantum mechanics, not a bug.’ Quantum probabilities are about indefiniteness, which is a feature of non-Booleanity just  as the nonlocal probabilistic correlations of entangled states\index{entanglement!entangled states!} are a feature of non-Booleanity. This has nothing to do with the size or complexity of a system---quantum probabilities about indefiniteness don’t become probabilities about ignorance just because a complex system is composed of a very large number of constituent elementary systems. 
\item `Although it can describe anything, a quantum description cannot include everything.’ Even though the Born probabilities refer to measurement outcomes in a macroscopic detector, a classical/Boolean system outside the quantum system under consideration,  this detector can always, in principle, be analyzed quantum mechanically, in which case the probabilities refer to a classical/Boolean system outside the new setup, and so on. The Heisenberg cut is movable---the world is not divided into a fixed classical part, not subject to quantum mechanical treatment, and a quantum part. In this sense, quantum mechanics is complete and objective. Nothing is left out of what the theory can in principle describe, and the description, although not a `view from nowhere,' is not subjective---the Born probabilities are objective kinematic\index{kinematic} features of the non-Boolean structure, `sui generis' and `there from the start'. 
\end{enumerate}
 
 From this standpoint, the onus is on rival interpretations to justify the appeal to many worlds as a response to indefiniteness, as in the Everett interpretation\index{Everett interpretation} \cite{Everett}, or to determinism, as in Bohm's theory\index{Bohm's theory} \cite{Bohm}, or to purely subjective probabilities as in QBism\index{QBism} \cite{deBrota}. For these interpretations, the salient feature of quantum mechanics is not non-Booleanity, and the theory is not about probability in the sense that special relativity is about space and time. Rather, it is about something else: reality as a branching structure of possible worlds coexisting within a universal wave function, or the causal structure of a deterministic world of  particles guided by a pilot wave, or an agent’s personal degrees of belief about possible experiences using the Born rule as a normative constraint to assign and update probabilities. The relevant question to ask is whether  these claims about the significance of the transition from classical to quantum mechanics are more plausible or more useful to physics than a neo-Bohrian\index{neo-Bohrian} analysis.

\section*{Acknowledgements}
Thanks  to Michel Janssen, Allen Stairs, and especially Michael Cuffaro and Jeremy Butterfield, for insightful and extremely helpful comments. Many thanks also to Krzysztof Sienicki who suggested several edits in a technical comment in arXiv: 2512.24198 on an earlier version of this paper.

\end{document}